\documentclass{acm_proc_article-sp}
\pdfoutput=1

\usepackage{amsmath, amsfonts, amssymb}
\usepackage{url}
\usepackage{graphicx}
\usepackage{xspace}

\def \insight {{\sffamily Insight}\xspace}
\def \impact {{\sffamily Impact}\xspace}
\def \libc {{\tt libc}\xspace}

\newcommand{\reffig}[1]{Fig.~\ref{#1}}
\newcommand{\component}[1]{{\sf #1}}
\newcommand{\xmltag}[1]{{\tt #1}}
\title{Simulating Cyber-Attacks for Fun and Profit}

\numberofauthors{4}

\author{
\alignauthor
Ariel Futoransky\\
       \affaddr{Corelabs}\\       
       \affaddr{Core Security Technologies}\\
       \email{futo@corest.com}        
\alignauthor
Fernando Miranda\\
       \affaddr{Corelabs}\\       
       \affaddr{Core Security Technologies}\\
       \email{fmiranda@corest.com}        
\alignauthor
Jos\'e Orlicki\\
       \affaddr{Corelabs}\\       
       \affaddr{Core Security Technologies}\\
       \affaddr{and ITBA}\\
       \email{jorlicki@corest.com}    
\and       
\alignauthor
Carlos Sarraute\\
       \affaddr{Corelabs}\\       
       \affaddr{Core Security Technologies}\\
       \affaddr{and ITBA}\\
       \email{carlos@corest.com}        
}

\begin{document}

\maketitle
\begin{abstract}

We introduce a new simulation platform called \insight, created to design 
and simulate cyber-attacks against large arbitrary target scenarios. \insight has surprisingly low hardware and configuration requirements, while making the simulation a realistic experience from the attacker's standpoint. The scenarios include a crowd of simulated actors: network devices, hardware devices, software applications, protocols, users, etc. 

A novel characteristic of this tool is to simulate vulnerabilities (including 0-days) and exploits, allowing an attacker to compromise machines and use them as pivoting stones to continue the attack. A user can test and modify complex scenarios, with several interconnected networks, where the attacker has no initial connectivity with the objective of the attack. 

We give a concise description of this new technology, and its possible uses in the security research field, such as pentesting training, study of the impact of 0-days vulnerabilities, evaluation of security countermeasures, and risk assessment tool.
\end{abstract}




\section{Introduction}

Computer security has become a necessity in most of today's computer uses and practices, 
however it is a wide topic and security issues can arise from almost everywhere: binary flaws 
(e.g., buffer overflows \cite{aleph96:smashing_stack}), 
Web flaws (e.g., SQL injection, remote file inclusion), protocol flaws (e.g., TCP/IP flaws \cite{bellovin89security}), 
not to mention hardware, human, cryptographic and other well known flaws.
 
Although it may seem obvious, it is useless to secure a network with a hundred firewalls 
if the computers behind it are vulnerable to client-side attacks. The protection provided by an 
{Intrusion Detection System} (IDS) is worthless against new vulnerabilities and 0-day attacks. 
As networks have grown in size, they implement a wider variety of more complex configurations 
and include new devices (e.g. embedded devices) and technologies. This has created new flows of 
information and control, and therefore new attack vectors. As a result, the job of both black hat
and white hat communities has become more difficult and challenging. 

The previous examples are just the tip of the iceberg, computer security is a complex field and
it has to be approached with a \emph{global} view, considering the \emph{whole} picture simultaneously: 
network devices, hardware devices, software applications, protocols, users, etcetera. 
With that goal in mind, we are going to introduce a new simulation platform called \insight, 
which has been created to design and simulate cyber-attacks against arbitrary target scenarios. 

In practice, the simulation of complex networks requires to resolve the tension between the
\emph{scalability} and \emph{accuracy} of the simulated subsystems, devices and data.
This is a complex issue, and to find a satisfying solution for this trade-off 
we have adopted the following design restrictions:
\begin{enumerate}
	\item Our goal is to have a simulator on a single desktop computer, running hundreds 
of simulated machines, with a simulated traffic realistic \emph{only} from the attacker's 
standpoint. 

\item Attacks within the simulator are \emph{not} launched by real attackers in the wild 
(e.g. script-kiddies, worms, black hats). As a consequence, the simulation does not have 
to handle exploiting details such as stack overflows or heap overflows. 
Instead, attacks are executed from an attack framework by \insight users who know 
they are playing in a simulated environment.
\end{enumerate}

To demonstrate our approach, \insight introduces a platform for executing 
attack experiments and tools for constructing these attacks.
By providing this ability, we show that its users are able to design 
and adapt attack-related technologies, and have better tests to 
assess their quality.
Attacks are executed from an attack framework which includes many 
information gathering and exploitation modules.
Modules can be scripted, modified or even added.

One of the major \insight features is the capability to simulate \emph{exploits}. 
An exploit is a piece of code that attempts to compromise a computer system via a specific vulnerability.
There are many ways to exploit security holes. If a computer programmer makes a programming 
mistake in a computer program, it is sometimes possible to circumvent security. 
Some common exploiting techniques are stack exploits, heap exploits,
format string exploits, etc. 

To simulate these techniques in detail is very expensive. 
The main problem is to maintain the complete state (e.g., memory, stack, heap, CPU registers) 
for every simulated machine. 
From the attacker's point of view, 
an exploit can be modeled as a magic string sent to a target machine to unleash a hidden feature 
(e.g., reading files remotely) with a probabilistic result. 
This is a lightweight approach, 
and we have sacrificed some of the realism in order to support     
very large and complex scenarios. For example, $1,000$ virtual machines and network devices 
(e.g., hubs, switches, IDS, firewalls) can be simulated on a single 
Windows desktop, each one running their own simulated OS, applications, vulnerabilities and file systems. 
Certainly, taking into account available technologies, it is not feasible to use a complete virtualization server (e.g., VMware) running thousands of images simultaneously. 

As a result, the main design concept of our implementation is 
to focus on the attacker's point of view, and \emph{to simulate on demand}. 
In particular, the simulator only generates information as requested by the attacker. By performing this 
on-demand processing, the main performance bottleneck comes from the ability of the attacker to request information 
from the scenario. Therefore, it is not necessary, for example, to simulate the complete TCP/IP packet traffic 
over the network if nobody is requesting that information. 
A more lightweight approach is to send data between network sockets writing in the memory 
address space of the peer socket, and leaving the full packet simulation as an option.

\section{Background \& related work}

Using simulated networks as a research tool to gather knowledge regarding the techniques, 
strategy and procedures 
of the black hat community is not a new issue. Solutions such as \emph{honeypots} and 
\emph{honeynets} \cite{honeynet:2006, spitzner02} were primarily designed to attract malicious 
network activities and to collect data. 

A precise definition for the term honeypot is given 
by The Honeynet Project \cite{honeynet04}:
\begin{quote}
	A honeypot is an information system resource whose value 
	lies in unauthorized or illicit use of that resource.
\end{quote}

Over the last decade a wide variety of honeypot systems were 
built \cite{honeypot03, Bailey05theinternet, song01, yegneswaran04design, Provos04avirtual}, 
both academic and commercial. Honeypots have emerged as an 
interesting tool for gathering knowledge on new methods used by attackers, 
and the underlying strength of the approach lies in its simplicity. Typically, 
honeypots offer minimal interaction with the attacker, emulating only small 
portions of a real network behavior. However, this simplicity is also a weakness: 
none of these systems execute kernel or application code that attackers seek 
to compromise, and only a few ones maintain a per-flow and per-protocol state to 
allow richer emulation capabilities. Thus, honeypots are most useful for capturing 
indiscriminate or large-scale attacks, such as worms, viruses or botnets, rather than 
very focused intrusions targeting a particular host \cite{potemkin05}.

In Table \ref{insightvshoneypot} we show the main differences
with our approach. In particular, we are interested 
in the ability to compromise machines, and use them as 
\emph{pivoting stones}\footnote{ In a network attack, to \emph{pivot} means to use a compromised machine as a stepping stone to reach further networks and machines, making use of its trust relationships. } to build complex multi-step attacks.

\begin{table}[h]
\begin{center}
\small
\begin{tabular}{ p{35mm}  p{35mm} }
\textbf{Honeypots-like tools} & \textbf{\insight} \\
	\hline\hline
		Design focus: to detect, understand and monitor real cyber-attacks. & 
		Design focus: to reproduce or mimic cyber-attacks, penetration test training, what-if and 0-day scenarios. \\
	\hline
		Attacks are launched by real attackers: worms, botnets, script-kiddies, black-hats. & 
		Attacks are launched by the \insight users: pentest and forensic auditors, security researchers. \\
	\hline
		Simulation up to transport layer. & 
		Simulation up to application layer, including vulnerabilities and exploits. \\
	\hline
		Stateless or (a kind of) per-flow and per-protocol state. & 
		Applications and machines internal state. \\
	\hline
		No exploit simulation. No pivoting support. & 
		Full exploit and agent (shellcode) simulation. Ability to pivot through a chain of 
		agents.\\
\end{tabular}
\end{center}
\caption{Honeypots vs. \insight.}
\label{insightvshoneypot}
\end{table}

\normalsize

In contrast, ``high interaction honeypots'' and virtualization technologies (e.g., VMware, Xen, Qemu) 
execute native system and application code, but the price of this 
fidelity is quite high. For example, the RINSE approach \cite{rinse05} is implemented over the iSSFNet network 
simulator, which runs on parallel machines to support real-time simulation of large-scale networks.
All these solutions share the same principle of simulating almost every aspect of a real machine or real network, but share the similar problems too: expensive configuration cost 
and expensive hardware and software licenses. 
Moreover, most of these solutions are not fully compatible with standard network protections (e.g., firewalls, IDSs), suffering a lack of integration between all security actors in complex cyber-attack scenarios.

\emph{Security assessment} and \emph{staging} are other well known security practices. It is common, for example in web application development, to duplicate the production environment on a staging environment (accurately mimicking or mirroring the first) to anticipate changes and their impact. The downside is: it is very difficult to adopt this approach in the case of network security due to several reasons. It would require the doubling of the hardware and software licenses and (among other reasons) there are no means to automatically configure the network.

Other interesting approaches to solve these problems include the framework developed by Bye et al. \cite{byeschmidt2008}. While they focus on distributed denial of service attacks (DDoS) and defensive IDS analysis, we focus on offensive strategies to understand the scenarios and develop countermeasures. Also Loddo et al. \cite{loddosaiu2008} have integrated \emph{User Mode Linux} \cite{dike2006} and \emph{Virtual Distributed Ethernet} \cite{davoli2005} to create a flexible and very detailed network laboratory and simulation tool. The latter project has privileged accuracy and virtualization over scalability and performance.

The \emph{Potemkin Virtual Honeyfarm} \cite{potemkin05} is another interesting prototype. 
It improves high-fidelity honeypot scalability by up to six times while still closely emulating the execution behavior of individual Internet hosts. Potemkin uses quite sophisticated on-demand techniques for instantiating 
hosts\footnote{Including \emph{copy-on-write} file system optimizations implemented also in \insight, as we are going to see it in \S\ref{filesystem}.}, but this approach focuses on attracting real attacks and it shows the same honeypot limitations to reach this goal. 
As an example, to capture e-mail viruses, a honeypot must posses an e-mail address, must be scripted to read mail (executing attachments like a naive user) and, most critically, \emph{real e-mail users} must be influenced to add the honeypot to their address books. 
Passive malware (e.g., many spyware applications) may require a honeypot to generate explicit requests, and focused malware (e.g., targeting only financial institutions) may carefully select its victims and never touch a large-scale honeyfarm. In each of these cases there are partial solutions, and they require careful engineering to truly mimic the target environment.

In conclusion, new trends in network technologies make cyber-attacks more difficult to understand, 
learn and reproduce, and the current tools to tackle these problems 
have some deficiencies when facing large complex scenarios. 
In spite of that, it is possible to overcome 
the problems described above using the lightweight software simulation tool we present.

\section{Insight approach \& overview}

\begin{figure}[th!]
\begin{center}
\includegraphics[scale =0.66]{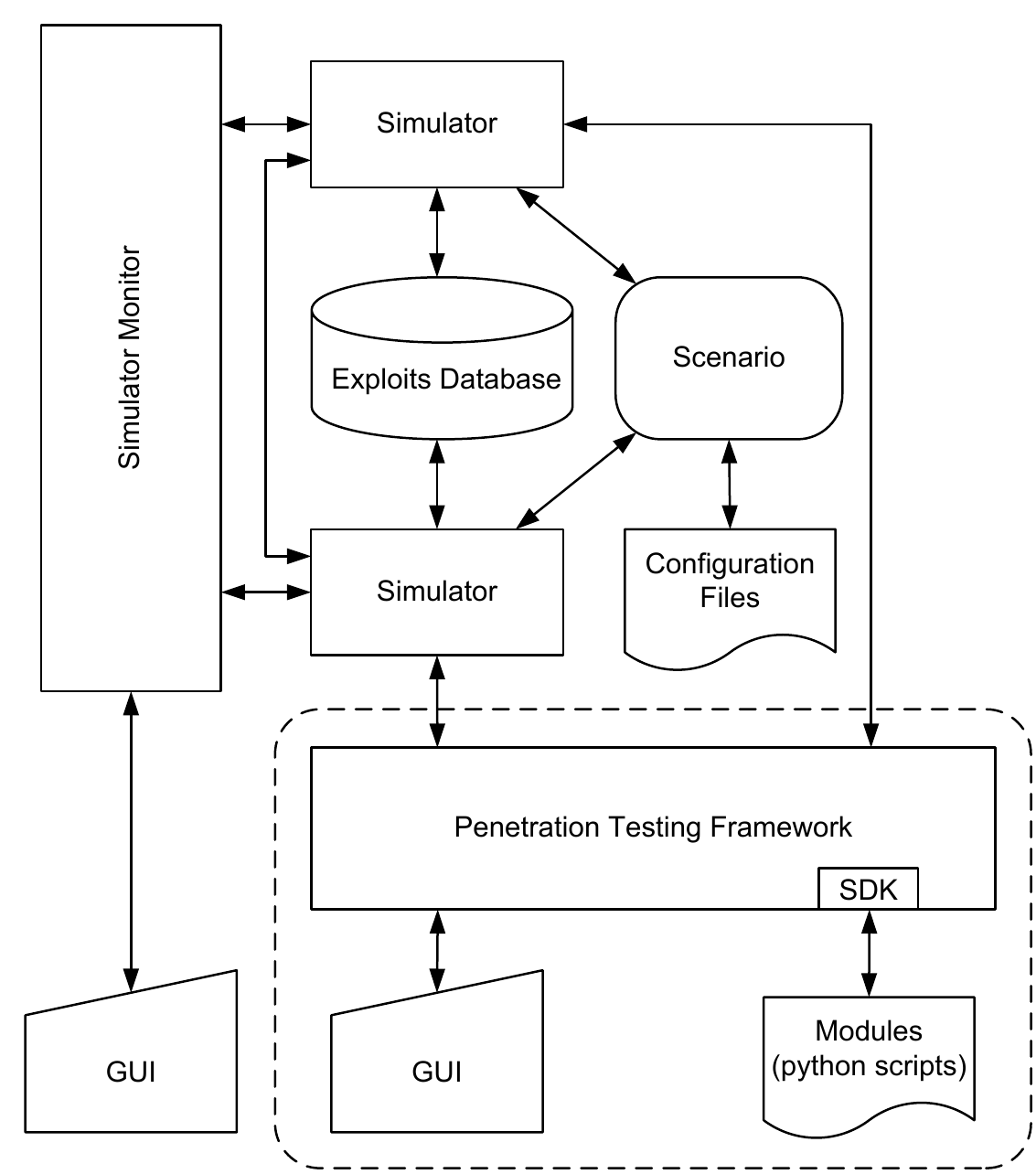}
\end{center}
\caption{\insight architecture layout.}
\label{fig:arch}
\end{figure}

A diagram of the \insight general architecture is showed in \reffig{fig:arch}. 
The \component{Simulator} subsystem is the main component. It performs all 
simulation tasks on the simulated machines, such as system call execution, 
memory management, interrupts, device I/O management, etcetera. 

At least one \component{Simulator} 
subsystem is required, but the architecture allows several ones, each running 
in a real computer (e.g., a Windows desktop). In this example, there are two simulation subsystems, 
but more could be added in order to support more virtual hosts. 

The simulation proceeds in a lightweight fashion. It means, for example, that 
not all system calls for all OS are supported by the simulation. Instead of 
implementing the whole universe of system calls, \insight handles a reduced and 
generic set of system calls, shared by all the simulated OS. Using this approach, 
a specific OS system call is mapped to an \insight syscall which works similarly 
to the original one. For example, the Windows sockets API is based on the 
Berkeley sockets API model used in Berkeley UNIX, but both implementations are 
slightly different\footnote{For additional details look at the Winsock API documentation 
(available from \url{http://msdn.microsoft.com}), which includes a section called Porting 
Socket Applications to Winsock.}. Similarly, 
there are some instances where \insight sockets have to diverge from strict adherence 
to the Berkeley conventions, usually due to implementation difficulties in the 
simulated environment. In spite of this (and ignoring the differences between OS), all 
sockets system calls of the real world have been mapped to this unique simulated API.

Of course, there are some system calls and management tasks closely 
related to the underlying OS which were not fully supported, such as UNIX {\sf fork} 
and {\sf signal} syscalls, or the complete set of functions implemented by the Windows SDK. 
There is a trade-off between precision and efficiency, and the decision of which syscalls 
were implemented was made with the objective of 
maintaining the precision of the simulation 
\emph{from the attacker's standpoint}.

The exploitation of binary vulnerabilities\footnote{\insight supports simulation for 
binary vulnerabilities. 
Other kind of vulnerabilities (e.g. client-side and SQL injections) will be implemented in the future versions.} 
is simulated with a \emph{probabilistic} 
approach
, keeping the attack model simple, lightweight, and avoiding to track 
anomalous conditions (and its countermeasures), such as buffer overflows, 
format string vulnerabilities, exception handler overwriting---among other well known vulnerabilities 
\cite{anley2007}. This probabilistic approach allows us to mimic 
the unpredictable behavior when an exploit is launched against a targeted machine.

Let us assume that a simulated computer was initialized with an 
underlying vulnerability (e.g. it hosts a vulnerable OS). 
In this case, the exploit payload is replaced by a 
special ID or ``magic string'', which is sent to the attacked 
application using a preexistent TCP communication channel. 
When the attacked application receives this ID, \insight will decide if the exploit 
worked or not based on a probability distribution that depends on the exploit 
and the properties describing the simulated computer (e.g., OS, patches, open services). 
If the exploit is successful, then \insight will grant the control 
in the target computer through the \emph{agent abstraction}, 
which will be described in \S \ref{sec:attack_model}.

The probabilistic attack model is implemented by the \component{Simulator} subsystems, 
and it is supported by the \component{Exploits Database}, a special configuration file 
which stores the information related to the vulnerabilities. This file has a XML tree structure, 
and each entry has the whole necessary information needed by the simulator to compute 
the probabilistic behavior of a given simulated exploit. For example, a given 
exploit succeeds against a clean XP SP2 with 83\% probability if port 21 is open, but 
crashes the system if it is a SP1. We are going to spend some time in the probability distribution, 
how to populate the exploits database, and the \insight attack model in the next sections.

Returning to the architecture layout showed in \reffig{fig:arch}, all simulator subsystems are coordinated 
by a unique \component{Simulator Monitor}, which deals with management and administrative operations, 
including administrative tasks (such as starting/stopping a simulator instance) and providing statistical 
information for the usage and performance of these.

A set of \component{Configuration Files} defines the \emph{snapshot} of a virtual 
\component{Scenario}. Similarly, a scenario snapshot defines the instantaneous status 
of the simulation, and involves a crowd of simulated actors: servers, workstations, 
applications, network devices (e.g. firewalls, routers or hubs) and their present status. 
Even users can be simulated using this approach, and this is especially interesting in 
client-side attack simulation, where we expect some careless users opening 
our poisoned crafted e-mails. 

Finally, at the right bottom of the architecture diagram, we can see the 
\component{Penetration Testing Framework}, an external system which interacts with 
the simulated scenario in real time, sending system call requests through a communication channel 
implemented by the simulator. This attack framework is a free tailored version of the 
\impact solution\footnote{Available from \url{http://trials.coresecurity.com/.}}, however 
other attack tools are planned to be supported in the future (e.g., Metasploit 
\cite{moore:cansecwest06}). 

The attacker actions are coded as \impact script files (using Python) 
called \emph{modules}, which have been implemented using the attack framework SDK, as shown 
in the architecture diagram. 
The framework Python modules include several tools for common tasks (e.g. information 
gathering, exploits, import scenarios). The attacks are executed in real 
time against a given simulated scenario; a simulation component can provide scenarios of thousands 
of computers with arbitrary configurations and topologies.  
\insight users can design new scenarios and they have scripts to manage the creation and modifications for the 
simulated components, and therefore iterate, import and reproduce cyber-attack experiments.

\section{The simulated attack model}\label{sec:attack_model}

One of the characteristics that distinguish the scenarios simulated by \insight is the ability to compromise machines, and use them as pivoting stones to build complex multi-step attacks.
To compromise a machine means to install an {agent} that will be able to execute arbitrary system calls (syscalls) as a user of this system.

The agent architecture is based on the solution called \emph{syscall proxy} 
(see \cite{corelabs:2002} for more details).
The idea of syscall proxying is to build a sort of \emph{universal payload} that allows an attacker
to execute any system call on a compromised host.
By installing a small payload (a thin syscall server) on a vulnerable machine,
the attacker will be able to execute complex applications on his local host,
with all system calls executed remotely. This syscall server is called an \emph{agent}.

In the \insight attack model, the use of syscall proxying introduces two additional layers between a process run by the attacker and the compromised OS. These layers are the \emph{syscall client} layer and the \emph{syscall server} layer.

The syscall client layer runs on the attacker's \component{Penetration Testing Framework}.
It acts as a link between the process running on the attacker's machine and the system services on a remote host simulated by \insight. 
This layer is responsible for forwarding each syscall argument and generating a proper request that the agent can understand. It is also responsible for sending this request to the agent and sending back the results to the calling process.

The syscall server layer (i.e. the agent that runs on the simulated system)
receives requests from the syscall client to execute specific syscalls using the OS services. After the syscall finishes, its results are marshalled and sent back to 
the client.

\subsection{Probabilistic exploits}

In the simulator security model, a vulnerability 
is a mechanism used to access an otherwise restricted communication channel. 
In this model, a real exploit payload is replaced by an ID or ``magic string'' 
which is sent to a simulated application. If this application 
is defined to be vulnerable (and some other requirements are fulfilled), then an agent 
will be installed in the computer hosting the vulnerable application.


The simulated exploit payload includes the aforementioned magic string.  
When the \component{Simulator} subsystem receives this information, it looks up for 
the string in the \component{Exploits Database}. If it is found, then the simulator will decide 
if the exploit worked or not and with what effect based on a probability distribution that depends on 
the effective scenario information of that computer and the specific exploit. Suppose, 
for example, that the \component{Penetration Testing Framework} assumes (wrongly) the attacked machine 
is a Red Hat Linux 8.0, but that machine is indeed a Windows system. In this hypothetical situation, 
the exploit would fail with 100\% of probability. On the other side, if the attacked 
machine is effectively running an affected version of Red Hat Linux 9.0, then the probability of success could be 75\%,
or as determined in the exploit database.

\subsection{Remote attack model overview}

\begin{figure}[ht!]
\begin{center}
\includegraphics[scale=0.60]{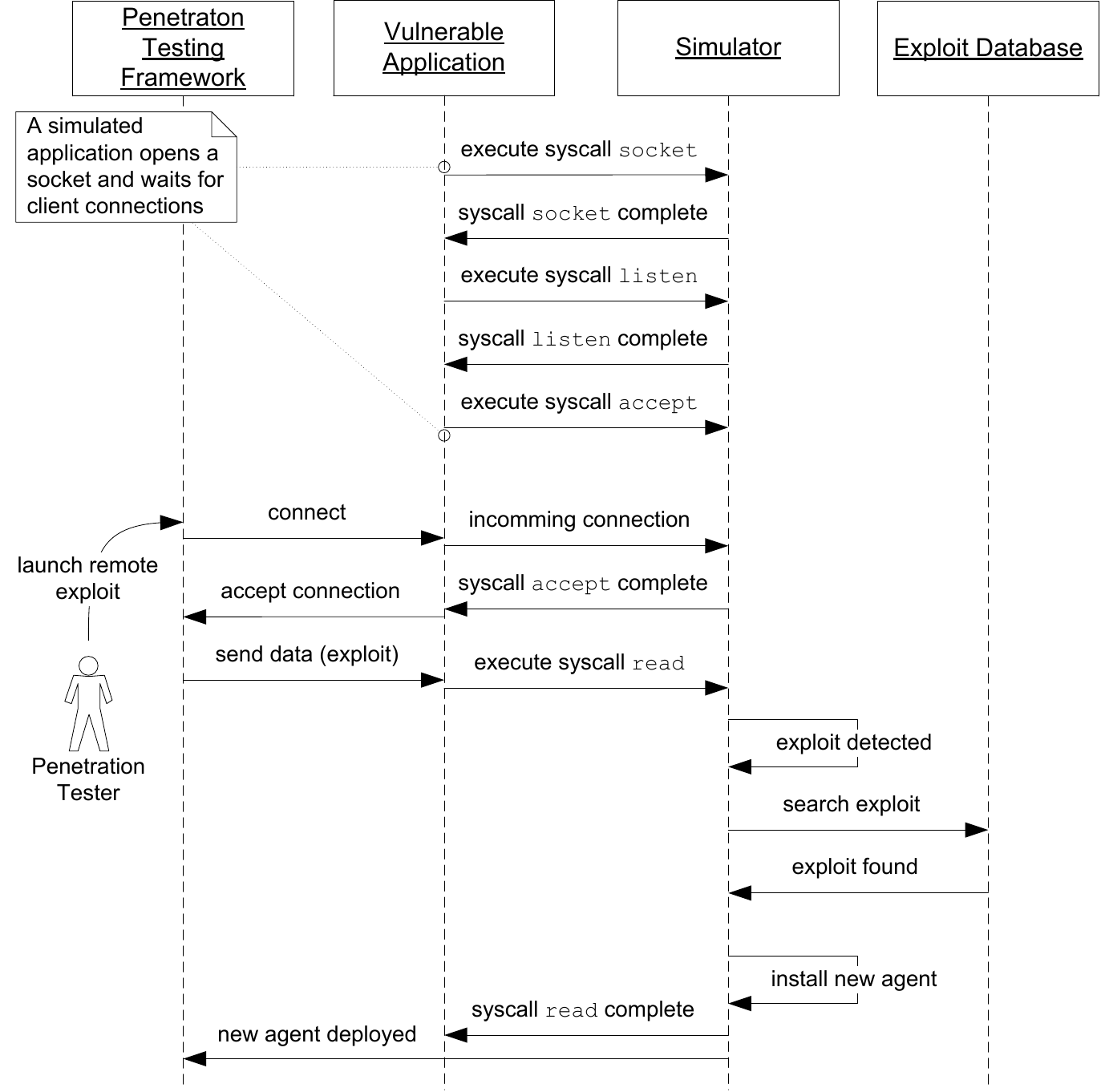}
\end{center}
\caption{Remote attack model.}
\label{fig:remote_attack_model}
\end{figure}

In \reffig{fig:remote_attack_model} we can see the sequence of events which occurs when an attacker 
launches a remote exploit against a simulated machine. The rectangles 
in the top are the four principal components involved: The \component{Penetration Testing Framework}, 
the \component{Simulator} and the \component{Exploits Database} are the subsystems explained in \reffig{fig:arch}; 
the \component{Vulnerable Application} is a simulated application or service which is 
running inside an \insight scenario and has an open port. 
In the diagram the declared components are represented as named rectangles, 
messages are represented as solid-line arrows, and time is represented as a vertical 
progression.

When an exploit is launched against a service running in a simulated machine, 
a connection is established between the \component{Penetration Testing Framework} 
and the service\footnote{This connection is established, for example, by a real Windows socket 
or a simulated TCP/IP socket, see \S\ref{st:sockets}.}. 
Then, the simulated exploit payload is sent to the application. 
The targeted application reads the payload by running the system call {\sf read}.
Every time the syscall {\sf read} is invoked, the \component{Simulator} subsystem analyzes if a 
magic string is present in the data which has just been read. When a magic string 
is detected, the \component{Simulator} searches for it in the \component{Exploits Database}. 
If the exploit is found, a new agent is installed in the compromised machine. 

The exploit payload also includes information of the OS that the 
\component{Penetration Testing Framework} knows about the attacked machine: OS version, system architecture, service packs, 
etcetera. All this information is used to compute the probabilistic function and allows the 
\component{Simulator} to decide whether the exploit should succeed or not. 

\subsection{Local attack model overview}

\insight can also simulate local attacks: If an attacker gains control over a machine 
but does not have enough privileges to complete a specific action, a local attack 
can deploy a new agent with higher privileges.  

\begin{figure}[h!]
\begin{center}
\includegraphics[scale=0.60]{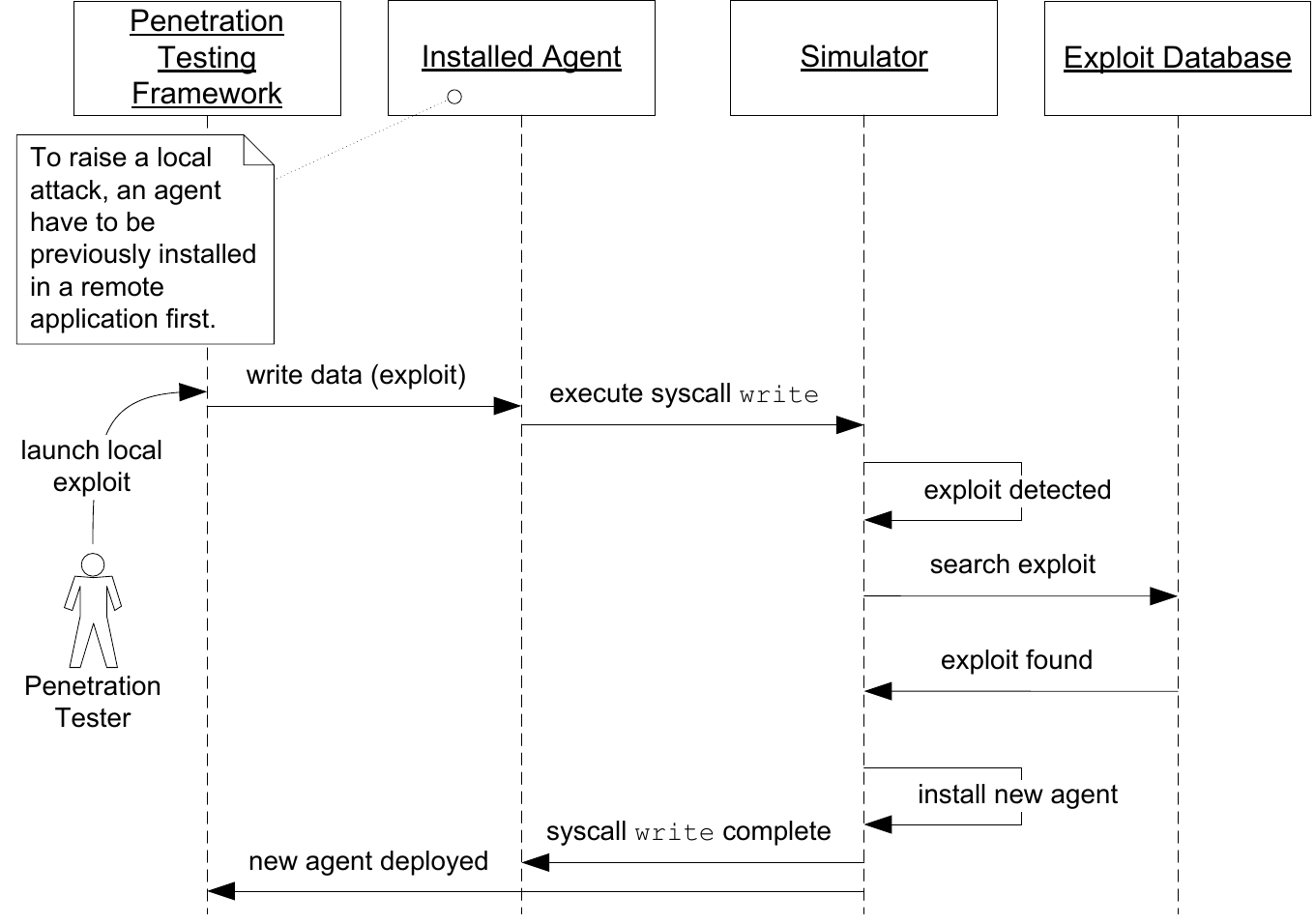}
\end{center}
\caption{Local attack model.}
\label{fig:local_attack_model}
\end{figure}

In \reffig{fig:local_attack_model} we can see the sequence of events which occurs when a local attack is launched against a given  machine. A running agent has to be present in the targeted machine in order to launch a local exploit. All local simulated attacks are executed by the \component{Simulator} subsystem identically: The \component{Penetration Testing Framework} will write the exploit magic string into the agent standard input, using the {\sf write} system call, and the \component{Simulator} will eventually detect the magic string intercepting that system call.

In a similar way as the previous example, the exploit magic string is searched in the database and a new agent (with higher privileges) is installed with probabilistic chance.

\section{Detailed description}

One of the most challenging issues in the \insight architecture is to resolve the tension between realism and performance. The goal was to have a simulator on a single desktop computer, running hundreds of simulated machines, with a simulated traffic realistic from a penetration test point of view. But there is a trade-off between realism and performance and we are going to discuss some of these problems and other architecture details in the following sections. 

\subsection{The Insight development library}

New applications can be developed for the simulation platform using a minimal 
\emph{C standard library}, a standardized collection 
of header files and library routines used to implement common operations such as: 
input, output and string handling in the C programming language.
 
This library---a partial \libc---implements the most common functions (e.g., read, write, open), allowing 
any developer to implement his own services with the usual compilers and 
development tools (e.g., gcc, g++, MS Visual Studio). For example, a web server could be implemented, linked with the 
provided \libc and plugged within the \insight simulated scenarios.

The provided \libc supports the most common system calls, but it is still incomplete 
and we were unable to compile complex open source applications. In spite of this, 
some services (e.g., a small DNS) and network tools (e.g., ipconfig, netstat) 
have been included in the simulation platform, and new system calls are planned to be 
supported in the future.

\subsection{Simulating sockets}
\label{st:sockets}
\begin{figure}[h!]
\begin{center}
\includegraphics[scale=0.68]{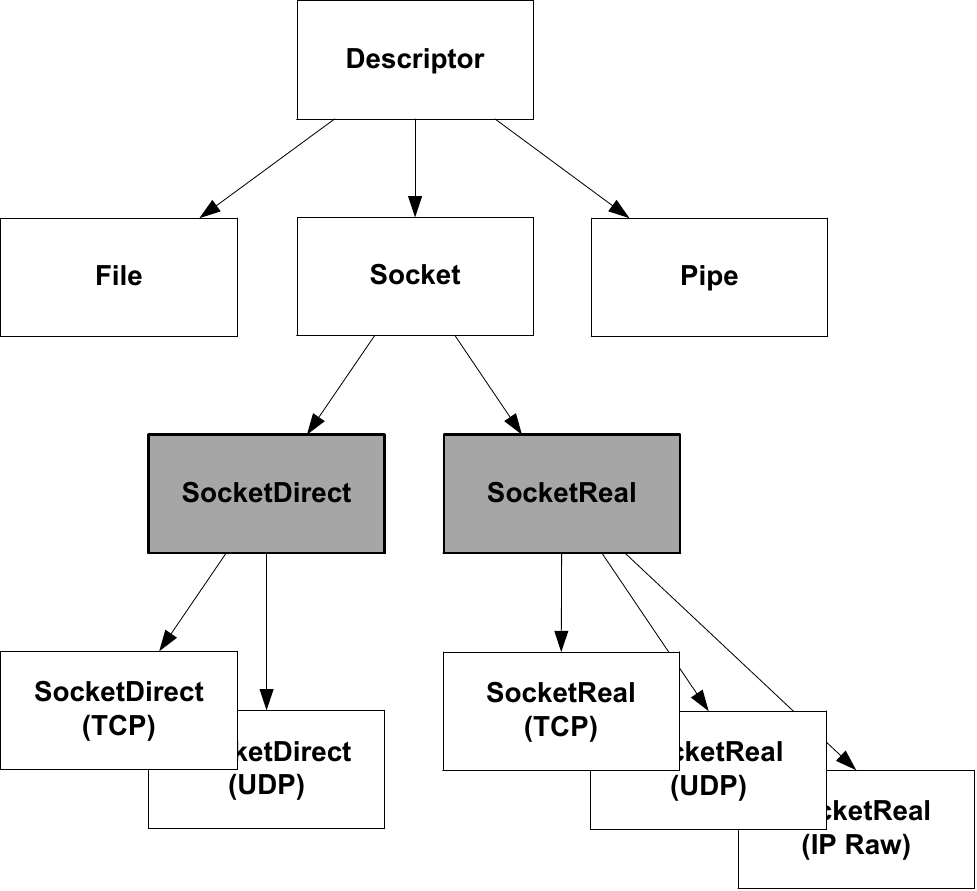}
\end{center}
\caption{Descriptors' object hierarchy tree.}
\label{fig:desc}
\end{figure}

A hierarchy for file descriptors has been developed as shown in \reffig{fig:desc}. 
File descriptors can refer (but they are not limited) to files, directories, sockets, or pipes. 
At the top of the hierarchy, the tree root shows the descriptor object which typically provides 
the operations for reading and writing data, closing and duplicating file descriptors, 
among other generic system calls.

The simulated sockets implementation spans between two kinds of 
supported sockets subclasses:
\begin{enumerate}
	\item SocketDirect. This variety of sockets is optimized for the simulation in one computer. 
	Socket direct is fast: as soon as a connection is established, the client keeps a file descriptor
	pointing directly to the server's descriptor. Routing is only executed during the connection and
	the protocol control blocks (PCBs) are created as expected, but they are only used during connection establishment.
	Reading and writing operations between direct sockets are carried out using shared memory. 
	Since both sockets can access the shared memory area like regular working memory, this is a very 
	fast way of communication.
	\item SocketReal. In some particular cases, we are interested in having full socket functionality. 
	For example, the communication between \insight and the outside world is made using real sockets. As a result, 
	this socket subclass wraps a real BSD socket of the underlying OS.
\end{enumerate}

Support for routing and state-less firewalling was also implemented, 
supporting the simulating of attack payloads that connect back to the attacker, 
accept connections from the attacker or reuse the attack connection.

\subsection{The exploits database}

When an exploit is raised, \insight has to decide whether the attack is successful 
or not \emph{depending on the environment conditions}. For example, an exploit can require 
either a specific service pack installed in the target machine to be successful, 
or a specific library loaded in memory, or a particular open port, among others requirements. 
All these conditions 
vary over the time, and they are basically unpredictable from the attacker's standpoint. 
As a result, the behavior of a given exploit has been modeled using a probabilistic approach.

In order to determine the resulting behavior of the attack, \insight uses the \component{Exploits Database} showed 
in the architecture layout of \reffig{fig:arch}. It has a XML tree structure. 
For example, if an exploit succeeds against a clean XP professional SP2 with 83\% probability, or crashes the machine with 0.05\% probability in other case; this could be expressed as follows:

\footnotesize
\begin{verbatim}
<database>
  <exploit id="sample exploit">
    <requirement type="system">
      <os arch="i386" name="windows" />
      <win>XP</win>
      <edition>professional</edition>
      <servicepack>2</servicepack>
    </requirement>
    <results>
      <agent chance="0.83" />   
      <crash chance="0.05" what="os" /> 
      <reset chance="0.00" what="os" /> 
      <crash chance="0.00" what="application" /> 
      <reset chance="0.00" what="application" /> 
    </results>
  </exploit>
  <exploit> ... </exploit>
  <exploit> ... </exploit>
  ...
</database>
\end{verbatim}
\normalsize

The conditions needed to install a new agent are described in the \xmltag{requirements} section. It is possible to use several tags in this section, they specify the conditions which have influence on the execution of the exploit (e.g., OS required, a specific application running, an open port). The \xmltag{results} section is a list of the relevant probabilities. 
In order, these are the chance of: 
\begin{enumerate}
	\item successfully installing an agent, 
	\item crashing the target machine, 
	\item resetting the target machine, 
	\item crashing the target application, 
	\item and the chance of resetting the target application.
\end{enumerate}

To determine the result, we follow this procedure: processing the lines in order, for each positive probability, 
choose a random value between 0 and 1. If the value is smaller than the chance attribute, 
the corresponding action is the result of the exploit.

In this example, we draw a random number to see if an agent is installed. 
If the value is smaller than 0.83, an agent is installed and the execution of the exploit is finished. 
Otherwise, we draw a second number to see if the OS crashes. 
If the value is smaller than 0.05, the OS crashes and the attacked machine becomes useless, 
otherwise there is no visible result.
Other possible results could be: raising an IDS alarm, writing some log in a network device (e.g. firewall, IDS or router) or capturing a session id, cookie, credential or password. 

The exploits database allows us to model the probabilistic behavior 
of any exploit from the attacker's point of view, but how do we populate our database? 
A paranoid approach would be to assign a probability of success of 100\% to every exploit. 
In that way, we would consider the case where an attacker 
can launch each exploit as many times as he wants, and will finally 
compromise the target machine with 100\% probability 
(assuming the attack does not crash the system).

A more realistic approach is to use statistics from real networks.
Currently we are using the framework presented by Marcelo Picorelli \cite{pico06} 
in order to populate the probabilities in the exploits database. This framework was 
originally implemented to assess and improve the quality of real exploits in QA environments. 
It allows us to perform over 500 real exploitation tests daily on several running configurations, 
spanning different target operating systems with their own setups and applications that 
add up to more than 160 OS configurations. In this context, a given exploit is executed against:
\begin{itemize}
	\item All the available platforms
	\item All the available applications
\end{itemize}

All these tests are executed automatically using low end hardware, VMware servers, OS images and snapshots. 
The testing framework has been designed to improve testing time and coverage, and we have modified it 
in order to collect statistical information of the exploitation test results.

\subsection{Scheduler}

The scheduler main task is to assign the CPU resources to the different 
simulated actors (e.g. simulated machines and process).
The scheduling iterates over the hierarchy machine-process-thread as a tree 
(like a depth-first search), each machine running its processes in round-robin.

In a similar way, running a process is giving all its threads the order to run until a system call 
is needed. Obviously, depending on the state of each thread, they run, 
change state or finish execution. The central issue is that threads execute systems calls and 
then (if possible) continue their activity until they finish or another system call is required. 

\insight threads are simulated within real threads of the underlying OS. 
Simulated machines and processes are all running within one or several working processes (running hundreds of threads), 
and all of them are coordinated by a unique scheduler process called the \emph{master process}.
Thanks to this architecture, there is a very low loss of performance due to context switching\footnote{Because 
descriptors and pointers remain valid when switching from one machine to the other.}.

\subsection{File system}
\label{filesystem}

In order to handle thousand of files without wasting huge disk space, the file system 
simulation is accomplished by mounting shared file repositories. 
We are going to refer these repositories as \emph{template file systems}. 
For example, all simulated Windows XP systems could share a file repository with 
the default installation provided by Microsoft. These shared templates would have 
reading permission only. Thus, if a virtual machine needs to read or change a file, it 
will be copied within the local file system of the given machine.  

This technique is well known as \emph{copy-on-write}. The fundamental idea is allowing multiple 
callers asking for resources which are initially indistinguishable, giving them pointers 
to the same resource. This function can be maintained until a caller tries to modify its copy 
of the resource, at which point a true private copy is created to prevent the changes from 
becoming visible to everyone else. All of this happens transparently to the callers. 
The primary advantage is that no private copy needs to be created if a caller never makes any modification.

On the other hand, with the purpose of improving the simulator's performance, a file cache has been implemented: 
the simulator saves the most recent accessed files (or block of files) in memory. In high scale 
simulated scenarios, it is very common to have several machines doing the same task at (almost) 
the same time\footnote{For example, when the simulation starts up, all UNIX machines would read 
the boot script from {\sf /etc/initd} file.}. If the data requested by these kind of tasks are in the 
file system cache, the whole system performance would improve, because less disk accesses will be required, 
even in scenarios of hundreds or thousands simulated machines.

\section{Performance analysis}

To evaluate the performance of the simulator we run a test including a scenario with
an increasing number of complete LANs with 250 computers each, simultaneously emulated. The tests only involves the
execution of a network discovery on the complete LANs through a TCP connection to port 80. 
An original pen-testing module used for information was executed with no modifications, this was 
a design goal of the simulator, to use real unmodified attack modules when possible.

\begin{table}[ht]
\small
\begin{center}
\begin{tabular}{c c c c}
\multicolumn{4}{c}{{\bf Performance of the simulator }} \\
\hline
\hline
LANs & Computers & Time (secs) & Syscalls/sec \\
\hline
1 & 250 & 80 & 356 \\
2 & 500 & 173 & 236 \\
3 & 750 & 305 & 175 \\
4 & 1000 & 479 & 139 \\
\end{tabular}
\end{center}
\caption{Evolution of the system performance as the simulated scenario grows, running a network discovery module, connecting to a predefined port. This benchmark was run on a single Intel Pentium D 2.67Ghz, 1.43GB RAM.}
\label{tab:performance}
\end{table}

\normalsize
We can observe the decrease of system calls processed per second as we increase the number of simulated computer as \insight was ran on a single real computer with limited resources. Nevertheless, the simulation is efficient because system calls are required on demand by the connections of the module gathering the information of the networks through TCP connections.

\section{Applications} 

We have created a playground to experiment with cyber-attack scenarios which has
several applications. The most important are:

\begin{description}
\item[Data collection and visualization.] Having the complete network scenario 
in one computer allows an easy capture and log of system calls and network traffic. 
This information is useful for analyzing and debugging real pen-test tools and their 
behavior on complex scenarios. Some efforts have been made to visualize attack 
pivoting and network information gathering using the platform presented.

\item [Pentest training.]

Our simulation tool is already being used in Pentest courses.
It provides reproducible scenarios, where students can practice the different steps of a pentest: information gathering, attack and penetrate, privilege escalation, local information gathering and pivoting.

The simulation allows the student to grasp the essence of pivoting.
Setting up a real laboratory where pivoting makes sense is an expensive task,
whereas our tool requires only one computer per student
(and in case of network / computer crash, the simulation environment can be easily reset).
Configuring new scenarios, with more machines or more complex topologies, is easy
as a scenario wizard is provided.

In Pentest classes with \insight, the teacher can check the logs to see if students used the right tools with the correct parameters.
He can test the students' ability to plan, see if they did not perform unnecessary actions.
The teacher can also identify their weaknesses as pentesters and plan new exercises to work on these.
The students can be evaluated: success, performance, stealth and quality of reports can be measured.

\item [Worm Spreading Analysis.] The lightweight design of the platform allows 
the simulation of socket/network behavior of thousands of computers gives a good 
framework for research on worm infestation and spreading. It should be possible to develop 
very accurate applications to mimic worm behavior using the \insight \texttt{C} 
programming API. There are available abstract modeling  \cite{chen:infocom2003} or high-fidelity 
discrete event \cite{songjie2005}
studies
but no system call level recreation of attacks like we propose in this future application of the platform.	

\item [Attack Planning.] It can be used as a flexible environment to develop and test 
attack planning algorithms used in automated penetration testing
based on attack graphs \cite{corelabs:2003}.

\item [Analysis of countermeasures.] Duplication of the production configuration on 
a simulated staging environment accurately mimicking or mirroring the security aspects 
of an organization's network allows the anticipation of software/hardware changes and 
their impact on security. For example, you can answer questions like ``Will the network 
avoid attack vector $A$ if firewall rule $R$ is added to the complex rule set $S$  
of firewall $F$?''

\item [Impact of 0-day vulnerabilities.]

The simulator can be used to study the impact of 0-days (vulnerabilities that
have not been publicly disclosed) in your network. 
How is that possible?  We do not know current 0-days...
but we can model the existence of 0-day vulnerabilities based on statistics.
In our security model, the specific details of the vulnerability 
are not needed to study the impact on the network, just that it may exist with 
a measurable probability.

That information can be gathered from public vulnerability databases:
the discovery date, exploit date, disclosure date and patch date
are found in several public databases of vulnerabilities and exploits
\cite{cert, securityfocus, secunia, frsirt}.

The risk of a 0-day vulnerability is given by the probability of an attacker discovering and exploiting it. Although we do not have data about the security underground,
the probabilities given by public information are a lower bound indicator.

\begin{figure}[h]
\begin{center}
\includegraphics[angle=0, width=0.40 \textwidth]{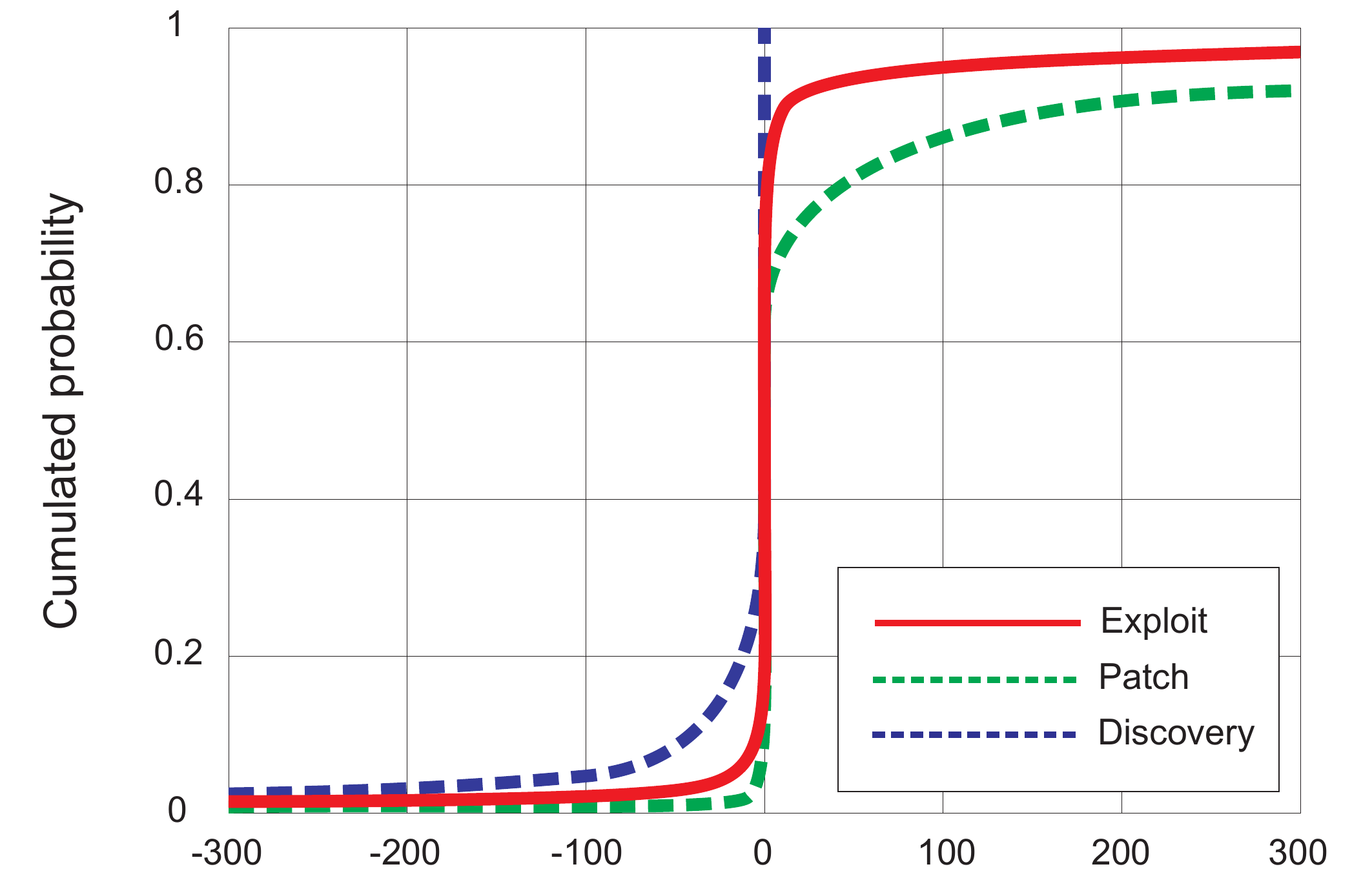}
\end{center}
\caption{Probabilities before disclosure.}
\label{fig:prob_disclosure}
\end{figure}

As shown in \cite{frei06}, the risk posed by a vulnerability 
exists before the discovery date,
augments as an exploit is made available for the vulnerability,
and when the vulnerability is disclosed.
The risk only diminishes as a patch becomes available 
and users apply the patches (and workarounds).

The probability of discovery, and the probability of an exploit being developed, 
can be estimated as a function of the time before disclosure
(see \reffig{fig:prob_disclosure} taken from \cite{frei06}).
For Microsoft products, we have visibility of upcoming disclosures of vulnerabilities:
every month (on patch Tuesday) on average 9,40 patches are released (high and medium risk),
based on those dates we estimate the probability that the vulnerabilities 
were discovered and exploited during the months before disclosure.

\end{description}

\section{Conclusion}

We have created a playground to experiment with cyber-attack scenarios.
The framework is based on a probabilistic attack model---that model is 
also used by attack planning tools developed in our lab.
By making use of the proxy syscalls technology, and simulating multiplatform agents,
we were able to implement a simulation that is both realistic and 
lightweight, allowing the simulation of networks with thousands of hosts.

The framework provides a global view of the scenarios.
It is centered on the attacker's point of view,
and designed to increase the size and complexity of simulated scenarios,
while remaining realistic for the attacker.

The value of this framework is given by its multiple applications:
\begin{itemize}
	\item Evaluate network security
	\item Evaluate security countermeasures
	\item Anticipate the risk posed by 0-day vulnerabilities
	\item Pentest training
	\item Worm spreading analysis
	\item Systematic study of Planning techniques
	\item Data generation to test visualization techniques
\end{itemize}

If you are interested in using \insight, send us an email.
We are trying to build a community 
using it as common language for discussing information security scenarios
and practices, and will strongly support new applications of this tool.

\bibliographystyle{plain}
\bibliography{bibdatabase}

\begin{thebibliography}{10}

\bibitem{anley2007}
Chris Anley, John Heasman, Felix Lindner, and Gerardo Richarte.
\newblock {\em The Shellcoder's Handbook}.
\newblock Wiley Press, 2nd edition, 2007.

\bibitem{Bailey05theinternet}
Michael Bailey, Evan Cooke, Farnam Jahanian, Jose Nazario, and David Watson.
\newblock The internet motion sensor: A distributed blackhole monitoring
  system.
\newblock In {\em In Proceedings of Network and Distributed System Security
  Symposium NDSS '05}, pages 167--179, 2005.

\bibitem{bellovin89security}
{S.}~{M.} Bellovin.
\newblock Security problems in the {TCP}/{IP} protocol suite.
\newblock {\em Computer Communications Review}, 19:2:32--48, 1989.

\bibitem{byeschmidt2008}
Rainer Bye, Stephan Schmidt, Katja Luther, and Sahin Albayrak.
\newblock Application-level simulation for network security.
\newblock In {\em Proceedings of the First International Conference on
  Simulation Tools and Techniques for Communications, Networks and Systems},
  2008.

\bibitem{corelabs:2002}
Maximiliano Caceres.
\newblock Syscall proxying - simulating remote execution.
\newblock Technical report, CoreLabs, Core Security Technology, 2002.
\newblock Available from \url{http://www.coresecurity.com}.

\bibitem{cert}
CERT.
\newblock {Computer Emergency Response Team, USA}.
\newblock \url{http://www.cert.org}.

\bibitem{chen:infocom2003}
{Z.} Chen, {L.} Gao, and {K.} Kwiat.
\newblock Modeling the spread of active worms.
\newblock In {\em Proceedings of IEEE INFOCOM 2003}, 2003.

\bibitem{davoli2005}
{R.} Davoli.
\newblock {VDE}: virtual distributed {Ethernet}.
\newblock In {\em First International Conference on Testbeds and Research
  Infrastructures for the Development of Networks and Communities}, volume~1,
  pages 213--220, 2005.

\bibitem{dike2006}
Jeff Dike.
\newblock {\em User Mode Linux}.
\newblock Prentice Hall, 1st edition, 2006.

\bibitem{frei06}
Stefan Frei, Martin May, Ulrich Fiedler, and Bernhard Plattner.
\newblock Large-scale vulnerability analysis.
\newblock In {\em LSAD '06: Proceedings of the 2006 SIGCOMM workshop on
  Large-scale attack defense}, pages 131--138, New York, NY, USA, 2006. ACM.

\bibitem{frsirt}
FrSirt.
\newblock {French Security Incident Response Team, France}.
\newblock \url{http://www.frsirt.com}.

\bibitem{corelabs:2003}
Ariel Futoransky, Luciano Notarfrancesco, Gerardo Richarte, and Carlos
  Sarraute.
\newblock Building computer network attacks.
\newblock Technical report, CoreLabs, Core Security Technology, 2003.
\newblock Available from \url{http://www.coresecurity.com}.

\bibitem{rinse05}
Michael Liljenstamand Jason Liuand David Nicoland Yougu Yuanand Guanhua
  Yanand~Chris Grier.
\newblock Rinse: The real-time immersive network simulation environment for
  network security exercises.
\newblock In {\em Workshop on Principles of Advanced and Distributed
  Simulation}, 2005.

\bibitem{loddosaiu2008}
Jean-Vincent Loddo and Luca Saiu.
\newblock Marionnet: A virtual network laboratory and simulation tool.
\newblock In {\em First International Conference on Simulation Tools and
  Techniques for Communications, Networks and Systems}, 2008.

\bibitem{honeypot03}
David Moore, Vern Paxson, Stefan Savage, Colleen Shannon, Stuart Staniford, and
  Nicholas Weaver.
\newblock Inside the slammer worm.
\newblock {\em IEEE Security and Privacy}, 1(4):33--39, 2003.

\bibitem{moore:cansecwest06}
{H.}~{D.} Moore.
\newblock Metasploitation.
\newblock In {\em CanSecWest 2006}, 2006.

\bibitem{aleph96:smashing_stack}
Aleph One.
\newblock Smashing the stack for fun and profit.
\newblock {\em Phrack}, 49--14, nov 1996.
\newblock Available from \url{http://www.phrack.com}.

\bibitem{pico06}
Marcelo Picorelli.
\newblock Virtualization in software development and {Q}{A}, 2006.
\newblock WMWORLD 2006 - \url{http://www.vmworld.com}.

\bibitem{honeynet04}
The~Honeynet Project.
\newblock {\em Know your enemy: Learning about security threats}.
\newblock Addison-Wesley Professional, 2nd edition, 2004.

\bibitem{honeynet:2006}
The~Honeynet Project.
\newblock Know your enemy: honeynets.
\newblock Technical report, Infocus At Securityfocus.com, May 2006.
\newblock \url{http://www.honeynet.org/papers/honeynet/}.

\bibitem{Provos04avirtual}
Niels Provos.
\newblock A virtual honeypot framework.
\newblock In {\em In Proceedings of the 13th USENIX Security Symposium}, pages
  1--14, 2004.

\bibitem{secunia}
Secunia.
\newblock \url{http://www.secunia.com}.

\bibitem{securityfocus}
SecurityFocus.
\newblock \url{http://www.securityfocus.com}.

\bibitem{song01}
{D.} Song, {R.} Malan, and {R.} Stone.
\newblock A snapshot of global internet worm activity.
\newblock Technical report, Arbor Networks, Nov 2001.

\bibitem{spitzner02}
{L.} Spitzner.
\newblock {\em Honeypots: Tracking Hackers}.
\newblock Addison-Wesley Longman Publishing Co., Inc., Boston, MA, USA, 2002.

\bibitem{potemkin05}
Michael Vrable, Justin Ma, Jay Chen, David Moore, Erik Vandekieft, Alex~{C.}
  Snoeren, Geoffrey~{M.} Voelker, and Stefan Savage.
\newblock Scalability, fidelity, and containment in the potemkin virtual
  honeyfarm.
\newblock {\em SIGOPS Oper. Syst. Rev.}, 39(5):148--162, 2005.

\bibitem{songjie2005}
Songjie Wei, Jelena Mirkovic, and Martin Swany.
\newblock Distributed worm simulation with a realistic internet model.
\newblock In {\em PADS '05: Proceedings of the 19th Workshop on Principles of
  Advanced and Distributed Simulation}, pages 71--79, Washington, DC, USA,
  2005. IEEE Computer Society.

\bibitem{yegneswaran04design}
{V.} Yegneswaran, {P.} Barford, and {D.} Plonka.
\newblock The design and use of internet sinks for network abuse monitoring.
\newblock In Proceedings of Recent Advances in Intrusion Detection (RAID),
  Sept. 2004.

\end{thebibliography}

\balancecolumns
\end{document}